\documentclass[a4paper,11pt]{article}

\usepackage{a4wide}
\usepackage{amsmath}

\title{Perturbations of
 a Schwarzschild black hole.\\
Notes on Zerilli's approach}

\author{Gianluca \textsc{Cruciani\footnote{cruciani@icra.it}}\\
{\small \textit{ICRA-International Centre for
Relativistic Astrophysics}}\\ {\small \textit{I-00185 Rome
(Italy)}}}

\date{December 8, 2005}

\begin{document}

\maketitle

\begin{flushleft}\small
PACS: 0420C\\
Ref.: {\it Il Nuovo Cimento B}, {\bf 120}, n. 10-11, 1045-1053
(2005).
\end{flushleft}\normalsize

\begin{abstract}
Modelling the free fall (and radiative phenomenology) of a massive
particle, charged or not, in a static and spherically symmetric
black hole is a classic, good relativistic dare that produced a
remarkable series of papers, mainly in the seventies of the past
century. Some formal topics about the mathematical machinery
required to perform the task are unfortunately still not very
clear; however, with the help of modern computer algebra
techniques, some results can at least be tested and corrected.
\end{abstract}

\section{Introduction}
When a massive particle falls in a black hole, it induces changes
in the spacetime metric that, although considered as an
``ordinary'' perturbation, can produce a phenomenology hardly
comparable to what it would be in a flat background. This fact,
even if dealing with elementary concepts of General Relativity,
found its first serious applications in a celebrated paper of 1957
by T. Regge and J. A. Wheeler \cite{rw} that somehow founded the
theory of the stability of the Schwarzschild black hole. However,
due to hard complications in the calculations, this investigation
came to a first end, after a number of intermediate stages (in
particular \cite{mathews, edelvish, zerilli1, zerilli2}), only in
1970-1974 with the work of F. J. Zerilli, who, in collaboration
with Wheeler and R. Ruffini \cite{jorufzer1, jorufzer2}, attacked
the problem of considering the perturbation of the e.m. and
gravitational fields produced in the falling in greater detail. In
\cite{zerilli3}, in particular, a realistic treatment is made,
considering the first-order stress-energy tensor contributions of
the perturbing particle, but the ideas lying at the base of the
angular decoupling of the equations still remain cryptical in some
of their aspects; it is now possible, also by the help of quite
popular computer algebra means ({\it Mathematica} by S. Wolfram
and the tensor calculus-dedicated package {\it MathTensor} by L.
Parker and S. Christensen), to account for some hidden aspects of
the logic path followed by the authors and fix some errors.

\section{Formulating the problem}

The perturbation analysis is performed by writing the Einstein's
system (in geometric units: $G=c=1$) as
\begin{equation}\label{eins}
G_{\mu\nu}(\mathbf{h})=8\,\pi\,T_{\mu\nu}(\mathbf{h})
\end{equation}
where $\mathbf{h}$ is the perturbation tensor accounting for the
presence of a massive particle, eventually endowed with an
electric charge, that acts directly on the spacetime geometry in
addition to the Schwarzschild metric tensor:
$\,g_{\mu\nu}=g^{\mbox{\bf \tiny S}}_{\mu\nu}+h_{\mu\nu}\,$ where
$\,g^{\mbox{\bf \tiny S}}_{\mu\nu}\,$ corresponds to the line
element
$$\,ds^2=-e^{\nu(r)}dt^2+e^{\lambda(r)}dr^2+
r^2(d\theta^2+\sin^2\theta\,d\phi^2)$$ with
$\nu(r)=-\lambda(r)=\ln (1-2\,m/r)$, $m$ being the mass
of the BH.\\
Historically speaking, almost all the ideas developed to perform a
pure metric perturbation analysis ({\it i.e.} with $T_{\mu\nu}=0$)
in the original paper of 1957 were adopted by the successive
works, in particular the ``harmonic syntax'' of the angular terms
and the distinction between two different kinds of perturbations,
belonging to different choices among the angular operators that
generate the multipole expansion itself.

\section{Nature and aims of the angular bases}

To write the equations in a smart basis, accounting for the
angular properties of a static, spherically symmetric problem (by
this fact allowing the separation of the radial parts), a set of
harmonic ``objects'' is originally introduced by Regge-Wheeler, by
a mechanism of parity splitting and repeated derivations of the
scalar harmonics; those objects are seen to form a useful basis
for rank-2 tensors in the Euclidean 3-dim. space and can
successively be modified to form a basis of the Minkowski
spacetime. They are called ``tensor multipole'' ({\it TM}) basis.
Later, Frank J. Zerilli modified the original choice in two of the
ten elements and noticed that the way those tensors are
constructed must be consistent with the procedure adopted by Jon
Mathews \cite{mathews}, consisting in a chain of external products
of elements, starting from the spherical vector basis, aimed to
the forming of a set which, in its spatial Euclidean version,
transforms under the irreducible representations of
$SO(3)$. This set is called ``tensor harmonic'' ({\it TH}) basis.\\
The procedure to obtain the {\it TM} basis resides on the combined
action of the time and radial projection, ${\bf P_t}$, ${\bf
P_r}$, the covariant derivative $\boldsymbol{\nabla}$ and orbital
angular momentum $\mathbf{L}=-i\,{\bf r}\times\boldsymbol{\nabla}$
operators on the scalar harmonics, that turn out to be linear
combinations of the {\it TH}s. Exploring the relationships between
bases that originate from external products of unit vectors and
those derived from operator compositions, one is soon aware of the
different characteristics of those sets concerning parity and with
respect to scalar products in $\mathcal{E}_c\otimes\mathcal{E}_c$
and in $\mathcal{L}_2(\mathcal{S}^2)$ (respectively, the Euclidean
complexified tensor product space and the square-integrable
functions' Hilbert space on the unit 2-sphere). This topic was
formerly treated in a couple of papers (\cite{dm1, dm2}) that
unfortunately contain a certain amount of errors and misprints.
Some of those results (those, in particular, about the
construction of a new $TH$ basis of $\mathcal{E}_c^3$ with
improved features) are revisited and presented in a correct,
implementable form in Appendix A.

\section{Scalar, vector and tensor harmonics in $\mathcal{E}_c^3$}

To reach a satisfying definition of rank-2 {\it TH}s, it is
natural to hierarchically construct them from simpler objects
belonging to
the same family: scalar and vector harmonics.\\
The basis of the Hilbert space of the square-integrable complex
functions on the unit sphere embedded in $\mathcal{E}^3$, whose
elements possess the property of being eigenfunctions of
$\mathbf{L}^2$ and $L_z$, is indicated by the double-indexed
function $Y_{JM}(\theta,\phi)$ (the scalar spherical harmonics),
and its full expression normally adopted (after having made
explicit the Legendre polynomials that appear in it) is:
\begin{equation}\label{scalhar}
\begin{split}
Y_{JM}(\theta,\phi)=&\,(-1)^{M/2}\sqrt{\frac{2\,J+1}{4\pi}}\,
\sqrt{\frac{(J-|M|)!}{(J+|M|)!}}\;(\cos^2\theta-1)^{|M|/2}\cdot\\
&\cdot\left\{\frac{\partial^{|M|}}{\partial(\cos\theta)^{|M|}}
\left[\frac{1}{2^J\,J!}\,
\frac{\partial^J}{\partial(\cos\theta)^J}\,(\cos^2\theta-1)^J
\right]\right\}\,e^{i M\phi}
\end{split}
\end{equation}
in which the spherical coordinates are, conventionally, $\theta\in
[0,\pi]$ (the polar angle referred to the z-axis) and $\phi\in
[0,2\pi)$ (the azimuthal angle referred to the x-axis of a
rectangular cartesian frame).\\
In fact, it is of common experience that the practical problem of
separating the angular parts in a set of equations that must be
reduced to a pure radial form can be efficaciously treated once a
gauge choice is made that transforms those angular parts in
algebraic expressions depending only on spherical harmonics of
same $(J,\,M)$ and their derivatives (usually up to the second
order in $\theta$). All combinations of those objects can be
rewritten in terms of, at most, two different harmonics, in
$(J,\,M)$ and $(J-1,\,M)$, making use of well-known recurrent
relations between Legendre polynomials of different $J$, leading
to the following formulas:
\begin{equation}\label{dty}
\frac{\partial\,Y_{JM}}{\partial\,\theta}=J\,\cot\theta\,Y_{JM}-
l(J,\,M)\,\csc\theta\,Y_{J-1,\,M}
\end{equation}
\begin{equation}
\frac{\partial^2\,Y_{JM}}{\partial\,\theta^2}=\{J^2\,\cot^2\theta-
[J\,(J+1)-M^2]\,\csc^2\theta\}\,Y_{JM}+l(J,\,M)\,\cot\theta\,
\csc\theta\,Y_{J-1,\,M}
\end{equation}
with $l(J,\,M)=\sqrt{\frac{(2\,J+1)\,(J^2-M^2)}{2\,J-1}}$.\\[1mm]
Those two expressions can be joined together, once noticed that
$\frac{\partial^2 Y_{JM}}{\partial\phi^2}=-M^2\,Y_{JM}$, in a
second order partial differential equation satisfied by the
$Y_{JM}$:
\begin{equation}\label{rel}
\frac{\partial^2\,Y_{JM}}{\partial\theta^2}+\csc^2\theta\,
\frac{\partial^2\,Y_{JM}}{\partial\phi^2}+\cot\theta\,
\frac{\partial\,Y_{JM}}{\partial\theta}+J\,(J+1)\,Y_{JM}=0\ .
\end{equation}
This relation will be used later to write in a more compact form
the Zerilli's $TM$s.\\[3mm]
The second step on the way to the {\it TH}s was made by Blatt and
Weisskopf \cite{blwei}, who introduced a basis of the space of the
complex vector fields on $\mathcal{S}^2$, which is built as:
\begin{equation}
Y^l_{JM}(\theta,\phi)=\langle1,m,1,n\,|\,J,M\rangle
\;Y_{lm}(\theta,\phi)\;\mathbf{e}_n
\end{equation}
where $\mathbf{e}_i$ symbolizes the generic element of the basis
of the complexified Euclidean space $\mathcal{E}_c^3$ composed by
the simultaneous eigenvectors of the spin operators $\mathbf{S},\
S_z$ with $S=1$:
\begin{align}
&e_1=-(\hat{x}+i\;\hat{y})/\sqrt{2}\nonumber\\
&e_0=\hat{z}\nonumber\\
&e_{-1}=(\hat{x}-i\;\hat{y})/\sqrt{2}
\end{align}
and the symbol $\langle\,|\,\rangle$ is the
bracketed Dirac notation of a {\it Clebsch-Gordan coefficient}.\\
After this definition, the rank-2 tensor space
$\mathcal{E}_c^3\otimes \mathcal{E}_c^3$ can be provided a proper
basis by considering orthonormalized external products of the
$\mathbf{e}_i$; keeping the formalism of \cite{zerilli1}:
\begin{equation}
t^{\,j}_m=\sum_{\mu=-1}^1\;\langle 1,\mu,1,m-\mu\,|\,j, m
\rangle\;\mathbf{e}_{m-\mu}\otimes\mathbf{e}_\mu
\end{equation}
Now, the {\it harmonic} tensor spherical basis, spanning the space
of the finite-dimensional irreducible representations of $SO(3)$,
can be defined as:
\begin{equation}
Y^{\,j\;l}_{\,J\,M}\,(\theta,\,\phi)=\sum_{m=-j}^j \langle
L,M-m,j,m\,|\,J,M\rangle\;t^{\,j}_m\;Y_{l\,M\!-m}(\theta,\phi)
\end{equation}
The orthonormality in $\mathcal{L}_2^2(\mathcal{S}^2)$ of these
objects is guaranteed once a scalar product in this tensor space
is defined as (the overbar meaning complex conjugation):
\begin{equation}\label{norm}
(T,S)=\int T:S\;d\Omega\equiv
\int_0^{\,2\pi}\hspace{-2mm}\int_0^{\,\pi}
\overline{T}^{\;\rho\sigma}S_{\rho\sigma}\,\sin\theta\,d\theta\,d\phi\
.
\end{equation}
It is worth noting that such a frame {\it is not} orthogonal in
$\mathcal{E}_c^3\otimes \mathcal{E}_c^3$ endowed with the scalar
product $(T,S)=T:S$, for fixed values of $J,\,M$.

\section{Tensor multipoles}

It is not difficult to directly write Zerilli's {\it TM} covariant
basis in $\mathcal{M}^4$ once a proper definition of $\mathbf{L}$
acting on scalar functions is adopted:
\begin{equation}
\mathbf{L}\,f=-i\,\mathbf{r}\wedge\mathbf{\nabla}\,f
\stackrel{\mbox{\tiny comp}}{=}-i\,r\,{E_\mu}^\rho\,f_{\,;\,\rho}
\end{equation}
where ${E_\mu}^\nu$ belongs to a rank-2 component of the
well-known Levi-Civita's tensor $\epsilon$:
\begin{equation}
{E_\mu}^\nu=\eta_{\,\mu\rho}\;{\epsilon}^{\,\rho\nu 01}=\left(
\begin{array}{cccc}
    0&0&0&0 \\
    0&0&0&0 \\
    0&0&0&\sin \theta \\
    0&0&-\frac{1}{\sin \theta} &0
\end{array}\right)\ ,
\end{equation}
$\eta_{\,\mu\nu}$ being the covariant Minkowski metric in
spherical coordinates. With this specification, the {\it TM}s are
defined as follows (the symbol ``${}_{()}$'' means that only the
symmetric part of the tensor is considered):
\begin{eqnarray}
&& \\[1mm]
&& \left.
\begin{array}{l}
TM_1=({\bf P_t}\circ {\bf P_t})\,Y_{JM}\\[1mm]
TM_2=({\bf P_r}\circ {\bf P_r})\,Y_{JM}\\[1mm]
TM_3=\sqrt{2}\,(i\,{\bf P_t}\circ {\bf P_r})_{()}\,Y_{JM}
\end{array}
\right\}\mbox{scalar components}\nonumber\\[2mm]
&& \left.
\begin{array}{l}
TM_4=m(J)\,(i\,r\,{\bf
P_t}\circ\boldsymbol{\nabla})_{()}\,Y_{JM}\\[1mm]
TM_5=m(J)\,(r\,{\bf P_r}\circ\boldsymbol{\nabla})_{()}\,Y_{JM}
\end{array}
\right\}\mbox{vector electric components}\nonumber\\[2mm]
&& \left.
\begin{array}{l}
TM_6=m(J)\,(-i\,{\bf
P_t}\circ \mathbf{L})_{()}\,Y_{JM}\\[1mm]
TM_7=m(J)\,({\bf P_r}\circ \mathbf{L})_{()}\,Y_{JM}
\end{array}
\right\}\mbox{vector magnetic components}\nonumber\\[2mm]
&& \hspace{2.1mm}TM_8=n(J)\left[({\bf P_r}\circ
\mathbf{L})_{()}+r\,(\mathbf{L}\circ\boldsymbol{\nabla})_{()}\right]\,
Y_{JM}\hspace{5mm}\mbox{tensor magnetic component}\nonumber\\[2mm]
&& \left.
\begin{array}{l}
TM_9=\frac{n(J)}{2}\left[(\mathbf{L}\circ
\mathbf{L})_{()}+3\,r\,({\bf P_r}\circ
\boldsymbol{\nabla})_{()}+r^2\,\boldsymbol{\nabla}\circ
\boldsymbol{\nabla}\right]\,Y_{JM}\\[1mm]
TM_{10}=\frac{m^2(J)}{2\,\sqrt{2}}\left[(\mathbf{L}\circ
\mathbf{L})_{()}-r\,({\bf P_r}\circ
\boldsymbol{\nabla})_{()}-r^2\,\boldsymbol{\nabla}\circ
\boldsymbol{\nabla}\right]\,Y_{JM}\nonumber
\end{array}
\right\}\mbox{tensor electric components}
\end{eqnarray}
where $m(J)=\sqrt{\frac{2}{J\,(J+1)}}$,
$n(J)=\sqrt{\frac{2}{(J-1)\,J\,(J+1)\,(J+2)}}$.\\
The explicit matrix form, with rows and columns labelled as
$(t,\,r,\,\theta,\,\phi)$, of these objects, with the help of
(\ref{rel}), can be cast as:\small
$$
\begin{array}{c}
TM_1=\left(\begin{array}{cccc}
Y_{JM}&0&0&0\vspace{2mm}\\ 0&0&0&0\vspace{2mm}\\ 0&0&0&0\vspace{2mm}\\
0&0&0&0
\end{array}\right)
TM_2=\left(\begin{array}{cccc}
0&0&0&0\vspace{2mm}\\ 0&Y_{JM}&0&0\vspace{2mm}\\ 0&0&0&0\vspace{2mm}\\
0&0&0&0
\end{array}\right)
TM_3=\left(\begin{array}{cccc} 0&\frac{i\,Y_{JM}}{\sqrt{2}}&0&0
\vspace{2mm}\\ *&0&0&0\vspace{2mm}\\
0&0&0&0\vspace{2mm}\\ 0&0&0&0
\end{array}\right)\vspace{5mm}\\
TM_4=\left(\begin{array}{cccc} 0&0&i\,r\,U&i\,r\,V\vspace{2mm}\\
0&0&0&0\vspace{2mm}\\ *&0&0&0\vspace{2mm}\\ *&0&0&0
\end{array}\right)\hspace{2mm}
TM_5=\left(\begin{array}{cccc} 0&0&0&0\vspace{2mm}\\
0&0&r\,U&r\,V \vspace{2mm}\\
0&*&0&0\vspace{2mm}\\ 0&*&0&0
\end{array}\right)\vspace{5mm}\\
TM_6=\left(\begin{array}{cccc} 0&0&\frac{r}{\sin\theta}\,V
&-r\,\sin\theta\,U\vspace{2mm}\\
0&0&0&0\vspace{2mm}\\ *&0&0&0\vspace{2mm}\\ *&0&0&0
\end{array}\right)\hspace{2mm}
TM_7=\left(\begin{array}{cccc} 0&0&0&0\vspace{2mm}\\
0&0&\frac{i\,r}{\sin\theta}\,V&-i\,r\,\sin\theta\,U\vspace{2mm}\\
0&*&0&0\vspace{2mm}\\0&*&0&0
\end{array}\right)\vspace{5mm}\\
TM_8=\left(\begin{array}{cccc} 0&0&0&0\vspace{2mm}\\
0&0&0&0\vspace{2mm}\\0&0&\frac{i\,r^2}{\sin\theta}\,X&-i\,
r^2\sin\theta\,W\vspace{2mm}\\0&0&*&-i\,r^2\sin\theta\,X
\end{array}\right)\hspace{2mm}
TM_9=\left(\begin{array}{cccc} 0&0&0&0\vspace{2mm}\\
0&0&0&0\vspace{2mm}\\0&0&r^2 W&r^2 X\vspace{2mm}\\
0&0&*&-r^2\sin^2\theta\,W\vspace{2mm}
\end{array}\right)
\end{array}$$
$$
\begin{array}{c}
TM_{10}=\left(\begin{array}{cccc} 0&0&0&0\vspace{2mm}\\
0&0&0&0\vspace{2mm}\\ 0&0&\frac{r^2 Y_{JM}}{\sqrt{2}}&0\vspace{2mm}\\
0&0&0&\frac{r^2\sin^2\theta\,\,Y_{JM}}{\sqrt{2}}\vspace{2mm}
\end{array}\right)
\hspace{4mm}\mbox{with}\hspace{0.8mm}\left\{\begin{array}{l}
U=\frac{m(J)}{2}\,\frac{\partial\,Y_{JM}}{\partial \theta}\\[2.5mm]
V=\frac{m(J)}{2}\,\frac{\partial\,Y_{JM}}{\partial
\phi}\\[2.5mm]
X=n(J)\left[\frac{\partial}{\partial\theta}-
\cot\theta\right]\frac{\partial\,Y_{JM}}{\partial \phi}\\[2mm]
W=n(J)\left[\frac{\partial^2}{\partial\theta^2}+\frac{J\,
(J+1)}{2}\right]Y_{JM}
\end{array}\right.
\end{array}
$$\normalsize\\[2mm]
where ``$\,*\,$'' denotes a symmetric component of a tensor.\\
Under a geometric point of view, the {\it TM}s can be classified
as:
\begin{itemize}
\item Space longitudinal ({\it i.e.} orthogonal to the unit
2-sphere) elements:\\$TM_2,\,TM_5,\,TM_7$ (in Zerilli's notation
``$a$'', ``$b$'', ``$c$''); \item Space transverse ({\it i.e.}
tangent to the unit 2-sphere) elements:\\
$TM_8,\,TM_9,\,TM_{10}$ (``$d$'',``$f$'',``$g$''); \item Time
addition elements:\\$TM_1,\,TM_3,\,TM_4,\,TM_6$
(``$a_0$'',``$a_1$'',``$b_0$'',``$c_0$'').
\end{itemize}
Being the ideas underlying the formation of this set mainly based
on the {\it euclidean} properties of transformation under {\bf L}
and $\boldsymbol{\nabla}$, orthonormalization problems were
expected to arise for the components belonging to the
$\mathcal{E}_c^3\rightarrow\mathcal{M}_c^4$ extension: in fact, it
is straightforward to see that the last three cited multipoles
($TM_{3,\,4,\,6}$) have their norm (defined by
(\ref{norm})) equal to -1.\\

\subsection{Regge-Wheeler perturbation tensors}

In the {\it TM} basis, ${\bf h}^m,\;{\bf h}^e$, the well-known
Regge-Wheeler perturbation tensors \cite{rw} that split
(\ref{eins}) in {\it magnetic} and {\it electric} parts, totally
decoupled from each other, can be written as:
\begin{eqnarray}
&& \\
&& {\bf h}^m =
\frac{2}{m(J)\,r}\,\big[-h_0(t,r)\,TM_6+i\,h_1(t,r)
\,TM_7\big]-\frac{i}{n(J)\,r^2}\,h_2(t,r)\,TM_8 \nonumber\\[3mm]
&& {\bf h}^e =
e^{\nu(r)}H_0(t,r)\,TM_1+e^{\lambda(r)}H_2(t,r)\,TM_2-i\,
\sqrt2\,H_1(t,r)\,TM_3 \nonumber\\
&& \hspace{8mm}+\frac{2}{m(J)\,r}\,\left[-i\,h_0(t,r)\,
TM_4+h_1(t,r)\,TM_5\right] \nonumber\\
&&
\hspace{8mm}+\frac{1}{n(J)}\,G(t,r)\,TM_9+\sqrt{2}\,\left[K(t,r)-\frac{1}{m^2(J)}\,
G(t,r)\right]\, TM_{10} \nonumber
\end{eqnarray}
and, once the celebrated gauge choices ($h_2=0$ for ${\bf h}^m$
and $h_0=h_1=G=0$ for ${\bf h}^e$) that share the same names are
applied, they reduce to:
\begin{eqnarray}
&& \\
&& {\bf h}^m = \frac{2}{m(J)\,r}\,\big[-h_0(t,r)\,TM_6+i\,h_1(t,r)
\,TM_7\big] \nonumber\\[3mm]
&& {\bf h}^e =
e^{\nu(r)}H_0(t,r)\,TM_1+e^{\lambda(r)}H_2(t,r)\,TM_2+
\sqrt2\,\left[-i\,H_1(t,r)\,TM_3+K(t,r)\,TM_{10}\right] \nonumber
\end{eqnarray}
making evident their belonging to different classes of
transformations of $SO(3)$.\\It is easier to explain their
decoupling in those terms than invoking an ``opposite parity''
feature which is hardly recognizable, once we refer to the
eigenvalues (if existing) of the operator $P$ that acts on a
function defined on $\mathcal{S}^2$ as $P(f(\theta,\,\phi))
=f(\pi-\theta,\,\phi+\pi)$: in the case of ${\bf h}^e$, it is
manifestly $P({\bf h}^e_{\mu\nu})=P(Y_{JM})=(-1)^J\,Y_{JM}$ for
any choice of $\mu,\,\nu$, while ${\bf h}^m$ has no defined
parity, since it is immediately seen from (\ref{dty}) that
$P(\partial_\theta Y_{JM})=(-1)^{J+1}\,Y_{JM}$ while
$\partial_\phi Y_{JM}=i\,M\,Y_{JM}$ so that $P(\partial_\phi
Y_{JM})=P(Y_{JM})=(-1)^J\,Y_{JM}$.

\appendix

\section{Harmonics and multipoles in the euclidean case}

In 1976 M. Daumens and P. Minnaert (Universit\'e de
Bordeaux-France) produced two papers \cite{dm1, dm2} about the
relationship between tensor harmonics and tensor multipoles in
both the Euclidean and the Minkowski space. Concerning the
euclidean 3-dim. space, to stress the fact that no conceptual
contradictions arise using one basis instead of the other in
separating the angular dependance of systems like (\ref{eins}),
once a unique definition of the quantum index J is adopted, they
derived an alternative $TH$ basis which had the property of being
traceless in all its components but one and parity-defined in the
traditional sense, contrary to Zerilli's $TM$s, and, contrary to
the formerly defined $TH$s, orthogonal also with respect to scalar
product in $\mathcal{E}_c^3\otimes\mathcal{E}_c^3$. Unfortunately,
some of the main formulas and relations concerning this topic
contain errors and/or misprints. The correct form of this harmonic
basis for rank-2 tensors in a Euclidean three-dimensional space is
presented here in a fully implementable form, illustrating the
linear combinations of the {\it TH}s and {\it TM}s previously
defined that generate
it.\\[3mm]
Called $\mathbf{v}$ the element of $\mathcal{S}_c^2$ whose
coordinates are
$\;\frac{(-1)^{j+l}}{\sqrt{3}}\,((-1)^{L+1},\,\,i,\,\,1)\;$, the
basis can be defined as:
\begin{equation}\label{ortbas}
X^{\,j\;l}_{\,J\,M}\,(\theta,\,\phi)=\sum_{L=|J-j|}^{J+j}\!\!\sqrt{3}\;
\mathbf{v}\cdot\mathbf{e}_{\mathrm{sign}(l)}\,\langle
j,|l|,J,-|l|\,|\,L,0\rangle\;Y^{\,j\;l}_{\,J\,M}\,(\theta,\,\phi)
\end{equation}
where $0\leq j\leq2$, $-j\leq l\leq j$ .\\
It is readily seen that those harmonics have the following parity:
\begin{equation}
X^{\,j\;l}_{\,J\,M}(-\vec{u})=\left\{{\rm sign}(l)+ \left[1-{\rm
sign}^2(l)\right](-1)^{j+J}\right\}\,X^{\,j\;l}_{\,J\,M}(\vec{u})\
,
\end{equation}
$\vec{u}$ being a generic element of $\mathcal{S}^2$. Every
multipole of order $j$ is found to be a linear combination of
tensor harmonics of the same order, following this scheme (that
corrects \cite{dm1} - Table 1):
\begin{eqnarray}
&& \\[1mm]
&& X^{\,0\;0}_{\,J\,M}=Y^{\,0\;J}_{\,J\,M} \nonumber\\
&& \left(
X^{\,1\;1}_{\,J\,M}\hspace{2mm}X^{\,1\;0}_{\,J\,M}\hspace{2mm}
X^{\,1\;-1}_{\,J\,M}\right)=\left(
Y^{\,1\;J+1}_{\,J\,M}\hspace{2mm} Y^{\,1\;J}_{\,J\,M}\hspace{2mm}
Y^{\,1\;J-1}_{\,J\,M}\right)\,A \nonumber\\
&&
\left(X^{\,2\;2}_{\,J\,M}\hspace{2mm}X^{\,2\;1}_{\,J\,M}\hspace{2mm}
X^{\,2\;0}_{\,J\,M}\hspace{2mm}X^{\,2\;-1}_{\,J\,M}\hspace{2mm}
X^{\,2\;-2}_{\,J\,M}\right)=\left(
Y^{\,2\;J+2}_{\,J\,M}\hspace{2mm}Y^{\,2\;J+1}_{\,J\,M}\hspace{2mm}
Y^{\,2\;J}_{\,J\,M}\hspace{2mm}Y^{\,2\;J-1}_{\,J\,M}\hspace{2mm}
Y^{\,2\;J-2}_{\,J\,M}\right)\,B  \nonumber
\end{eqnarray}
where $A$ and $B$ are two orthogonal matrices, confirming that the
(\ref{ortbas}) structure of a linear transformation with
Clebsch-Gordan coefficients effectively corresponds to a {\it
rotation}:
\begin{equation}
A=\left(
\begin{array}{ccc}
0 & -\sqrt{\frac{J+1}{2J+1}} & \sqrt{\frac{J}{2J+1}}
\vspace{2mm}\\
1 & 0 & 0\vspace{2mm}\\
0 & \sqrt{\frac{J}{2J+1}} & \sqrt{\frac{J+1}{2J+1}}
\end{array}\right)
\end{equation}
\begin{equation}
B=\left(
\begin{array}{ccccc}
\sqrt{\frac{J(J-1)}{2(2J+1)(2J+3)}} &
-\sqrt{\frac{2J(J+2)}{(2J+1)(2J+3)}} &
\sqrt{\frac{3(J+1)(J+2)}{2(2J+1)(2J+3)}} & 0 & 0 \vspace{1mm}\\
0 & 0 & 0 & -\sqrt{\frac{J+2}{2J+1}} & \sqrt{\frac{J-1}{2J+1}}
\vspace{1mm}\\
\sqrt{\frac{3(J-1)(J+2)}{(2J-1)(2J+3)}} &
-\sqrt{\frac{3}{(2J-1)(2J+3)}} &
-\sqrt{\frac{J(J+1)}{(2J-1)(2J+3)}} & 0 & 0 \vspace{1mm}\\
0 & 0 & 0 & \sqrt{\frac{J-1}{2J+1}} & \sqrt{\frac{J+2}{2J+1}}
\vspace{1mm}\\
\sqrt{\frac{(J+1)(J+2)}{2(2J-1)(2J+1)}} &
\sqrt{\frac{2(J+1)(J-1)}{(2J-1)(2J+1)}} &
\sqrt{\frac{3J(J-1)}{2(2J-1)(2J+1)}} & 0 & 0
\end{array}\right)
\end{equation}\\
Finally, the linear combinations of those $TH$s that generate
Zerilli's $TM$s can be easily found by a simple matricial calculus
procedure:
\begin{itemize}
\item[a)] called $\mathcal{J}$ the jacobian matrix of the
transformation from the cartesian rectangular coordinates $(x, y,
z)$ to the polar spherical $(\theta, \phi)$ on $\mathcal{S}^2$:
\begin{equation}
\mathcal{J}=\left(
\begin{array}{ccc} \sin\theta\,\cos\phi &
\cos\theta\,\cos\phi &
-\frac{\sin\phi}{\sin\theta}\vspace{2mm}\\
\sin\theta\,\sin\phi & \cos\theta\,\sin\phi &
\frac{\cos\phi}{\sin\theta}\vspace{2mm}\\
\cos\theta & -\sin\theta & 0
\end{array}\right)
\end{equation}
and $\mathcal{M}$ the Euclidean 3-dim. covariant metric tensor:
\begin{equation}
\mathcal{M}=\left(
\begin{array}{ccc}
1 & 0 & 0\vspace{2mm}\\
0 & 1 & 0 \vspace{2mm}\\
0 & 0 & \sin^2\theta
\end{array}\right)
\end{equation}
the operator linking the spherical representation $T_s$ of a
tensor to its cartesian one, $T_c$, can be expressed, in the
matrix algebra terms, as:
\begin{equation}
T_s=\mathcal{O}_{cs}(T_c)=\mathcal{J}^{-1}\,T_c\,\mathcal{J}
\,\mathcal{M}\;
\end{equation}
\item[b)] called $TM^{(3)}_i$ the Euclidean 3-dim. tensors
introduced in \cite{zerilli1}, counterparts of Zerilli's
multipoles of the previous section (as matrices, they are obtained
by deleting the first row and column from the (2, 5, 7, 8, 9,
10)-labelled ones);
\end{itemize}
it is readily found that:
\begin{eqnarray}
&& \\
&& TM^{(3)}_1=\mbox{``$a^{(3)}$''}=-\frac{1}{\sqrt{3}}\;
\mathcal{O}_{cs}(X^{\,0\;0}_{\,J\,M})+\sqrt{\frac{2}{3}}\;
\mathcal{O}_{cs}(X^{\,2\;0}_{\,J\,M}) \nonumber\\
&& TM^{(3)}_2=\mbox{``$b^{(3)}$''}=\mathcal{O}_{cs}
(X^{\,2\;-1}_{\,J\,M}) \nonumber\\
&& TM^{(3)}_3=\mbox{``$c^{(3)}$''}=\mathcal{O}_{cs}
(X^{\,2\;1}_{\,J\,M}) \nonumber\\
&& TM^{(3)}_4=\mbox{``$d^{(3)}$''}=\mathcal{O}_{cs}
(X^{\,2\;2}_{\,J\,M}) \nonumber\\
&& TM^{(3)}_5=\mbox{``$f^{(3)}$''}=\mathcal{O}_{cs}
(X^{\,2\;-2}_{\,J\,M}) \nonumber\\
&&
TM^{(3)}_6=\mbox{``$g^{(3)}$''}=-\sqrt{\frac{2}{3}}\;\mathcal{O}_{cs}
(X^{\,0\;0}_{\,J\,M})-\frac{1}{\sqrt{3}}\;
\mathcal{O}_{cs}(X^{\,2\;0}_{\,J\,M}) \nonumber
\end{eqnarray}

\end{document}